\documentclass[%
reprint,
amsmath,amssymb,
 aps,
floatfix
]{revtex4-1}

\usepackage{graphicx}% Include figure files
\usepackage{dcolumn}% Align table columns on decimal point
\usepackage{bm}% bold math
\usepackage{braket}
\usepackage{enumerate}
\usepackage{rotating}
\usepackage{comment}
\usepackage{tikz}
\usepackage{pgfplots}
\usepackage{appendix}

\usepackage{xcolor}
\usepackage{multirow}

\usepackage{amssymb}
\usepackage{bm}
\usepackage[normalem]{ulem}
\def\jek#1{{\color{cyan}{#1}}}  % Jorgen
\begin{document}
%\preprint{APS/123-QED}

\title{Signs in isotope shifts: a perennial headache}
% Force line breaks with \\

%
\author{Michel Godefroid}
\email[]{mrgodef@ulb.ac.be}
\affiliation{Spectroscopy, Quantum Chemistry and Atmospheric Remote Sensing (SQUARES), CP160/09, Universit\'e libre de Bruxelles (ULB), 1050 Brussels, Belgium}

\author{Jörgen Ekman}
\email[]{jorgen.ekman@mau.se}
\affiliation{Department of Materials Science and Applied Mathematics, Malm\"o University, SE-20506 Malm\"o, Sweden} 

\author{Per J\"onsson}
\email[]{per.jonsson@mau.se}
\affiliation{Department of Materials Science and Applied Mathematics, Malm\"o University, SE-20506 Malm\"o, Sweden}

\date{\today}% It is always \today, today,

\begin{abstract}
Signs of the different contributions to the isotope shift of an atomic line are discussed in details to clarify some confusing  differences between the electronic parameters calculated with RIS~\cite{Nazetal:2013a,Ekmetal:2018a} and those appearing in other commonly used expressions.
\end{abstract}

\maketitle
%\tableofcontents

\bibliographystyle{apsrev4-1}

%\bibliographystyle{unsrt}

%\vspace*{2cm}
\section{RIS3 and RIS4 sign conventions}

The sign conventions used in the description of the isotopic shift RIS3~\cite{Nazetal:2013a} and RIS4~\cite{Ekmetal:2018a} of the {\sc Grasp} package~\cite{Grasp2018} are explained in details in section~3 of Ekman {\it et al.}~\cite{Ekmetal:2018a}. 
When evaluating the frequency isotope shift $\delta \nu_{k,\mathrm{IS}}$ of a given spectral line $k$ from the relevant electronic and nuclear factors, 
%However 
one should pay attention to the definition of the frequency isotope shift itself. The latter is defined in RIS~\cite{Nazetal:2013a,Ekmetal:2018a}  as
\begin{equation}
\label{eq:RIS_IS}
  [\delta \nu_{k,\mathrm{IS}}^{A,A'}]^{\mbox{{\tiny RIS}}}=
  \nu_k^A-\nu_k^{A'} \; ,
\end{equation}
 with the convention $A>A'$, while in many other works~\cite{Aufetal:87a,Cheetal:2012a,Blaetal:2013a}, it is defined the other way round, as in the original work compilation of Heilig and Steudel (HS)~\cite{HeiSte:74a}, i.e. 
\begin{equation}
\label{HS_RIS_Diff_nu}
  [\delta \nu_{k,\mathrm{IS}}^{A,A'}]^{\mbox{{\tiny HS}}}=
  \nu_k^{A'}-\nu_k^{A} = - [\delta \nu_{k,\mathrm{IS}}^{A,A'}]^{\mbox{{\tiny RIS}}} \; ,
\end{equation}
with the convention $A'>A$. These two different conventions, $A>A'$ for RIS, and $A'>A$ for HS, restore the  sign compatibility for the isotope shifts of a spectral line calculated with both definitions,  (\ref{eq:RIS_IS}) and  
(\ref{HS_RIS_Diff_nu}). According to Heilig and Steudel~\cite{HeiSte:74a} or Bauche and Champeau~\cite{BauCha:76a} indeed, the isotope shift of a spectral line is called  {\it positive when the heavier isotope is shifted towards larger wavenumbers},  i.e. $\nu_k^{A'} > \nu_k^{A}$ when $ A' > A$.
 Keeping the same convention for a {\it positive frequency IS} requires the condition $ A > A'$ if adopting the RIS definition~(\ref{eq:RIS_IS}). 
 Such positive isotope shifts correspond to the following cases
\[ \hspace*{-4.0cm}
%\label{HS_RIS_Relations}
  [\delta \nu_{k,\mathrm{IS}}^{A,A'}]^{\mbox{{\tiny HS}}}=
  - [\delta \nu_{k,\mathrm{IS}}^{A',A}]^{\mbox{{\tiny HS}}}
\]
   \begin{eqnarray}
\label{HS_RIS_Relations}
%  [\delta \nu_{k,\mathrm{IS}}^{A,A'}]^{\mbox{{\tiny HS}}}=
%  - [\delta \nu_{k,\mathrm{IS}}^{A',A}]^{\mbox{{\tiny HS}}}
 = - [\delta \nu_{k,\mathrm{IS}}^{A,A'}]^{\mbox{{\tiny RIS}}}
 = + [\delta \nu_{k,\mathrm{IS}}^{A',A}]^{\mbox{{\tiny RIS}}} 
 & > & 0  \hspace*{1cm}  (A' > A) \\
 & < & 0 \hspace*{1cm}  (A'< A)
  \end{eqnarray}
that can be resumed in the following table
 \begin{center}
 \begin{tabular}{ccc}
        &   
$ [\delta \nu_{k,\mathrm{IS}}^{A,A'}]^{\mbox{{\tiny HS}}}$ &
$ [\delta \nu_{k,\mathrm{IS}}^{A,A'}]^{\mbox{{\tiny RIS}}}$ \\
\hline
  $A' > A$~~~~ & $>$ 0 & $<$ 0 \\
 $A' < A$~~~~ & $<$ 0 & $>$ 0
 \end{tabular}
 \end{center}
taking the $A \leftrightarrow A'$ permutation property $\delta \nu^{A',A} = - \delta \nu^{A,A'}$  into account.

Note that the RIS definition (\ref{eq:RIS_IS}) is consistent with previous works~\cite{GodFro:99a,Bloetal:01a,Caretal:2010a,CarGod:2013a,Sietal:2021a} defining the isotope shift on the electron-affinity  
\begin{equation}
    \mbox{IS}(A,A') = \delta \; ^e \! A \equiv \; ^e \! A(A) - \; ^e \! A(A')\jek{\;.}
\end{equation}
as being positive for $A' < A$.
 
 \subsection{Mass shift}
The total  mass shift is often written as~\cite{Cheetal:2012a}
 \begin{equation}
\label{eq:MS_Heilig_Steudel}
  [\delta \nu_{k,\mathrm{MS}}^{A,A'}] ^{\mbox{{\tiny HS}}} =
  M_k \frac{A'-A}{AA'} \; ,
\end{equation}
while RIS defines it as
\begin{equation}
\label{RIS4_eq_12}
[\delta \nu_{k,\mathrm{MS}}^{A,A'}]^{\mbox{{\tiny RIS}}} =\left(\frac{M'-M}{MM'}\right)\frac{\Delta K_{\mathrm{MS}}}{h}
= \left(\frac{M'-M}{MM'}\right) \Delta \tilde{K}_{\mathrm{MS}} \; .
\end{equation}
 
The mass factor having the same sign in both (\ref{eq:MS_Heilig_Steudel}) and (\ref{RIS4_eq_12}), the sign difference (\ref{HS_RIS_Diff_nu}) forces us to accept that
\begin{equation}
    \label{K_eq_minus_M}
\Delta \tilde{K}_{\mathrm{MS}} = - M_k \; .
\end{equation}

If  the normal mass shift (NMS) is the only contribution to the mass isotope shift, the latter is necessarily positive, corresponding to a ``normal'' isotope shift when referring to the mass IS of lines in hydrogen-like (single-electron) system. In that case, we have
\begin{eqnarray}
\label{Sign_NMS}
M_k & > & 0 \; ,\\
\Delta \tilde{K}_{\mathrm{MS}}  & < & 0  \; .
\end{eqnarray}
For many-electron systems, the specific mass shift (SMS) due to the mass polarization term can counter-balance the normal mass shift (NMS). It can be even large enough to produce an ``anomalous" mass isotope shift~\cite{GodFro:99a,CarGod:2013a},  corresponding to a smaller frequency for the heavier isotope.
%i.e. a  negative $\delta \nu$ (keeping the choice A > A') ???

\subsection{Field shift}
The field shift (FS) is expressed in RIS as
\begin{equation}
\label{RIS3_FS_eq_24}
[\delta \nu_{k,\mathrm{FS}}^{A,A'}]^{\mbox{{\tiny RIS}}} 
= F_k \; [\delta \langle r^2 \rangle^{A,A'}]^{\mbox{{\tiny RIS}}}
\end{equation}
with
\begin{equation}
\label{RIS3_FS_eq_29} 
[\delta \langle r^2 \rangle^{A,A'}]^{\mbox{{\tiny RIS}}}
= \langle r^2 \rangle^{A} - \langle r^2 \rangle^{A'} \; .
\end{equation}
Unfortunately, the sign of the change in the root-mean-square nuclear charge radii (\ref{RIS3_FS_eq_29}) also differs with other (commonly used) conventions~\cite{HeiSte:74a,Aufetal:87a,Cheetal:2012a,Blaetal:2013a}
\begin{equation}
\label{HS_RIS_Diff_delta_r}
[\delta \langle r^2 \rangle^{A,A'}]^{\mbox{{\tiny RIS}}}
= - [\delta \langle r^2 \rangle^{A,A'}]^{\mbox{{\tiny HS}}} \; .
\end{equation}

\subsection{Total isotope shift}

Adding the mass shift (\ref{RIS4_eq_12}) and the field shift (\ref{RIS3_FS_eq_24}), the  total isotope shift calculated with the RIS codes writes as
\begin{equation}
\label{RIS4_total_IS}
[\delta \nu_{\mathrm{IS}}^{A,A'}]^{\mbox{{\tiny RIS}}} =\left(\frac{M'-M}{MM'}\right) 
[\Delta \tilde{K}_{\mathrm{MS}}]^{\mbox{{\tiny RIS}}}
+ F  [ \delta \langle r^2 \rangle^{A,A'}]^{\mbox{{\tiny RIS}}} \; .
\end{equation}

Applications of (\ref{RIS4_total_IS}) can be found in the multiconfiguration Dirac-Hartree-Fock calculations of electronic isotope shift factors in Li-like ions~\cite{Lietal:2012a},
Be-like ions~\cite{Nazetal:2014a}, Ba~I~\cite{Nazetal:2015a}, Mg~I~\cite{Filetal:2016b}, Al~I~\cite{Filetal:2016c}, Zn~I~\cite{Filetal:2017a}, Sb~I~\cite{Gametal:2018a}, Os~I~\cite{Paletal:2016a}, Ir~I~\cite{SchGod:2021a},
and of In$^-$, In~I, Tl$^-$ and Tl~I~\cite{Sietal:2021a}, using the auxiliary computational tool {\tt fical}  included in the {\sc Grasp} package.

Due to the sign differences (\ref{K_eq_minus_M}) and (\ref{HS_RIS_Diff_delta_r}), this last expression differs from  Cheal {\it et al.}~\cite{Cheetal:2012a}~(CCF)
\begin{equation}
\label{Cheal_total_IS}
 [\delta \nu_{\mathrm{IS}}^{A,A'}]^{\mbox{{\tiny CCF}}} = \left(\frac{A'-A}{AA'}\right) [M]^{\mbox{{\tiny CCF}}}
+ F  [ \delta \langle r^2 \rangle^{A,A'}]^{\mbox{{\tiny CCF}}}
\end{equation}
or from Blaum {\it et al.}~\cite{Blaetal:2013a} (BDN)
\begin{equation}
\label{Blaum_total_IS}
 [\delta \nu_{\mathrm{IS}}^{A,A'}]^{\mbox{{\tiny BDN}}} = \left(\frac{M_{A'}-M_{A}}{M_{A}M_{A'}}\right) 
 [K_{MS}]^{\mbox{{\tiny BDN}}}
+ F  [ \delta \langle r^2 \rangle^{A,A'}]^{\mbox{{\tiny BDN}}}
\end{equation}
who adopt the Heilig and Steudel definition of $\delta \langle r^2 \rangle^{A,A'}$, i.e. 
$[\delta \langle r^2 \rangle^{A,A'}]^{\mbox{{\tiny HS}}}
= \langle r^2 \rangle^{A'} - \langle r^2 \rangle^{A}$.
 Comparing (\ref{RIS4_total_IS}) with (\ref{Cheal_total_IS}) or 
(\ref{Blaum_total_IS}), and taking into account that
\begin{equation}
\label{Total_IS_Comparison}
[\delta \nu_{\mathrm{IS}}^{A,A'}]^{\mbox{{\tiny RIS}}}
= - [\delta \nu_{\mathrm{IS}}^{A,A'}] ^{\mbox{{\tiny CCF}}} 
= - [\delta \nu_{\mathrm{IS}}^{A,A'}] ^{\mbox{{\tiny BDN}}}  \; , 
\end{equation}
one deduces that the field shift factor $F$ is the same for all conventions while
\begin{equation}
\label{sign_inv}
 [\Delta \tilde{K}_{\mathrm{MS}}] ^{\mbox{{\tiny RIS}}}
= - [M] ^{\mbox{{\tiny CCF}}}
= -  [K_{MS}] ^{\mbox{{\tiny BDN}}}  \; .
\end{equation}
The {\it ab initio} calculated  electronic mass parameters,
$[\Delta \tilde{K}_{\mathrm{MS}}]^{\mbox{{\tiny RIS}}}$, had therefore to be sign-inverted according to (\ref{sign_inv}) to be properly treated in experimental studies adopting 
(\ref{Cheal_total_IS}) or (\ref{Blaum_total_IS}) expressions~\cite{Bisetal:2016a,Xieetal:19a,Muketal:2020a}.

\section{Recommendations}

To avoid misinterpretation or misuse of the electronic parameters estimated with atomic structure codes, it is highly recommended to emphasize the possibility of sign conflict between the different conventions. The minimal recommendation is to refer to the appropriate isotope shift expression, (\ref{RIS4_total_IS}), (\ref{Cheal_total_IS}) or (\ref{Blaum_total_IS}), with the relevant crucial sign conventions (\ref{eq:RIS_IS}) or (\ref{HS_RIS_Diff_nu}) used for $\delta \nu_{\mathrm{IS}}^{A,A'}$ and 
$\delta \langle r^2 \rangle^{A,A'}$.

\section{Helpful considerations}

For the FS contribution, we should keep in mind that:

\begin{enumerate}
\item Deeper is the potential, more bound are the electronic levels.
%\item 
%An electronic level is more bound if the potential is deeper.
\item
The finite volume effect eliminates the singularity of the point-charge nucleus model and therefore decreases the binding energy.
\item
This reduction of the binding energy increases with the electron density at the nucleus. Therefore, for a given isotope, omitting the mass shift,  the frequency of a line (always positive) is 
\item
\begin{enumerate}
\item
 {\it smaller}  for a finite nuclear charge than for a point-charge nucleus if the electronic density at the nucleus is {\it smaller} for the upper level than for the lower level,
\item
 {\it larger} for a finite nuclear charge than for a point-charge nucleus if the electronic density at the nucleus is {\it larger} for the upper level than for the lower level.
\end{enumerate}
\item
For a given isotopic pair and a given electronic transition, we should consider the sign of $\delta \langle r^2 \rangle$ when comparing the effect of the finite nuclear charge distribution between the two isotopes. Referring to the above  terminology, we can qualify the frequency shift due to the volume effect as a ``normal'' or ``anomalous'' field shift: 
\begin{enumerate}
\item 
if the heaviest isotope has a {\it larger} nuclear volume than the lightest isotope, (increasing of the nuclear radius with the number of nucleons), 
\begin{enumerate}
    \item 
the line frequency  will be {\it smaller} for the heaviest isotope if we are in case 4(a) ({\it decreasing}  of the electron density when exciting the atom) i.e., anomalous FS.
    \item
the line frequency  will be {\it larger} for the heaviest isotope if we are in case 4(b) ({\it increasing}  of the electron density when exciting the atom) i.e., normal FS.
\end{enumerate}
\item
if the heaviest isotope has a {\it smaller} nuclear volume than the lightest isotope, (decreasing of the nuclear radius with the number of nucleons), 
\begin{enumerate}
\item
the line frequency  will be {\it larger} for the heaviest isotope if we are in case 4(a) ({\it decreasing}  of the electron density when exciting the atom) i.e., normal FS.
\item
the line frequency will be {\it smaller} for the heaviest isotope if we are in case 4(b) ({\it increasing} of the electron density when exciting the atom) i.e., anomalous FS.
\end{enumerate}
\end{enumerate}
\end{enumerate}

\begin{acknowledgments}
\noindent 
The authors are grateful to Simon Lechner, Patrick Palmeri, Sacha Schiffmann and Ran Si for useful discussions.
\end{acknowledgments}

%\bibliography{atoms}
%

\end{document}